\documentclass[aps,pra,twocolumn,showpacs,float,superscriptaddress]{revtex4-1}
\usepackage{amsmath,amsthm,latexsym,amssymb,amsfonts,epsfig,graphicx} 
\usepackage[english]{babel}
\usepackage[cp1250]{inputenc}	
\usepackage{texdraw} 
\usepackage[margin=2.5cm]{geometry}	
\usepackage{color} 
\usepackage[T1]{fontenc}        
\usepackage[QX]{fontenc}        
\usepackage{bm}
\usepackage{lmodern}

\usepackage{epstopdf} 
\usepackage[usenames,dvipsnames,svgnames,table]{xcolor}
\usepackage{array}
\usepackage{float}
\usepackage{multirow}
\usepackage{setspace}
\usepackage{esint}
\usepackage{indentfirst}
\usepackage[colorlinks=true, citecolor=red, linkcolor=blue ]{hyperref}
\usepackage{natbib}

\newcommand{\be}{\begin{equation}}    
\newcommand{\ee}{\end{equation}}
\newcommand{\ba}{\begin{eqnarray}}
\newcommand{\ea}{\end{eqnarray}}

 \newcommand{\reff}[1]{(\ref{#1})}
 
\newcommand{\BR}{\mathbf{R}}

\newcommand{\Br}{\mathbf{r}}

\newcommand{\BF}{\mathbf{F}}
\newcommand{\BL}{\mathbf{L}}
\newcommand{\BJ}{\mathbf{J}}

\newcommand{\bee}{\begin{eqnarray}}
\newcommand{\eee}{\end{eqnarray}}
\newcommand{\m}{\mbox{\boldmath $\mu$}}

\begin{document}
\title{Two dipolar atoms in a harmonic trap}

\author{Rafał Ołdziejewski}
\affiliation{Center for Theoretical Physics, Polish Academy of Sciences, Al. Lotników 32/46, 02-668 Warsaw, Poland}

\author{Wojciech Górecki}
\affiliation{Center for Theoretical Physics, Polish Academy of Sciences, Al. Lotników 32/46, 02-668 Warsaw, Poland}

\author{Kazimierz Rzążewski}
\affiliation{Center for Theoretical Physics, Polish Academy of Sciences, Al. Lotników 32/46, 02-668 Warsaw, Poland}

\date{November 26, 2015}


\begin{abstract}
Two identical dipolar atoms moving in a harmonic trap without an external magnetic field are investigated. Using the algebra of angular momentum a semi - analytical solutions are found. We show that the internal spin - spin interactions between the atoms couple to the orbital angular momentum causing an analogue of Einstein - de Haas effect. We show a possibility of adiabatically pumping our system
from the s-wave to the d-wave relative motion. The effective spin-orbit coupling occurs at anti-crossings of the energy levels.
\end{abstract}

\maketitle
\section{Introduction} \label{Intro}
We observe a remarkable progress in experiments with ultra cold quantum
gases. Many are performed with large number of atoms in a single trap.
However, progress is also made at a level  of a few atoms in a trap. These
experiments are performed with cold atoms distributed between the wells of
an optical lattice. This way, with a help of tunable parameters of
interaction, using the Feshbach resonances~\cite{Ketterle1998,Courteille1998}, and the properties of
the lattice itself, one can access in a controlled  way vital models of
condense matter physics - for recent review see~\cite{lewenstein2012ultracold}. The realization of an original idea of R. Feynman~\cite{feynman1982} of building
quantum simulators is becoming realized. Quantum simulators are more akin 
to classical, special purpose analogue computers rather than their general,
all purpose digital cousins.
Many experiments in optical lattices are performed with the Mott insulator
phase~\cite{Greiner2002,Greiner2002a,Bakr2010} where a well defined, small number of atoms is
confined in every well. Another set of a few atoms
in a trap experiments is offered by the setting available in Selim Jochim's
lab~\cite{Serwane2011}. Detailed properties of such systems crucially
depend on the properties of atom-atom interaction. This interaction is
best tested if exactly two atoms are present. Early analytic predictions for contact interacting atoms~\cite{Busch1998}  where positively verified in precise spectroscopic experiments~\cite{Kohl2006}.
New twist to the problem is introduced by the long range dipole-dipole
(DD) interactions~\cite{Ospelkaus2006,Grishkevich2009}. Negligible in early days of quantum gases experiments,
dipole-dipole interactions are getting more and more relevant with the
condensation of chromium~\cite{Griesmaier2005,Stuhler2005}, erbium~\cite{Aikawa2012,Baier2015,Frisch2015} and recently dysprosium~\cite{Lu2011,Lu2012,Maier2014,Tang2015,Maier2015}. The dipolar interaction couples spin degree of freedom
with the orbital angular momentum. This leads to the well known Einstein -
de Haas effect~\cite{Einstein1915}. To observe this effect with chromium atoms, where DD
interaction is just a perturbation, properly resonant magnetic field
strength must be used~\cite{Gawryluk2007}. Of course a direct coupling to the orbital
angular momentum is possible for sufficiently strong DD interactions. For
the large systems it has been noted using conventional mean field approach~\cite{Kawaguchi2006}.
It is the purpose of this paper to present exact analysis of the role of
DD interactions for two atoms trapped in a spherically symmetric harmonic
potential. In Section 2 we introduce our theoretical model. The simplicity
of the harmonic potential allows to separate the center of mass degree of
freedom. The relative motion hamiltonian remains spherically symmetric.
Utilizing this symmetry we may construct the energy eigenstates using the
angular momentum algebra. What remains is the set of coupled radial
Schr\"{o}dinger equations linking components of the wave function 
corresponding to orbital angular momenta differing by two units. We model
the radial component of the interaction by the dipolar expression modified
by the infinite repulsive sphere at short distances. In Section 3 we
present our results. We note the anti-crossings of the energy levels as a
function of the dipolar coupling constant. This dependence may be tuned by the change of the trapping frequency. The most striking
feature is the possibility of adiabatically pumping our system from the
s-wave to the d-wave relative motion.
\section{Theoretical model} \label{Theomod}

Let us consider two identical dipolar atoms (fermions or bosons) of a spin (a total angular momentum of an atom) $f_{1}=f_{2}$ moving in an isotropic harmonic trap. The Hamiltonian of such a system can be written as:

\be H=-\frac{1}{2}\nabla_{1}^{2}-\frac{1}{2}\nabla_{2}^{2}+\frac{1}{2}r_{1}^{2}+\frac{1}{2}r_{2}^{2}+V(\Br_{1}-\Br_{2})\label{hamiltoniancaly}\ee

where $\Br_{1}$ and $\Br_{2}$ are the position vectors of the two atoms. We are using harmonic oscillator units, in which $\hbar \omega$ is an unit of energy and the characteristic size of the ground state of the trap $\sqrt{\frac{\hbar}{m\omega}}$ is a length unit. An interaction potential $V(\Br_{1}-\Br_{2})$ is a sum of a short range Van der Waals (VdW) and a long range magnetic dipole - dipole interaction potentials. We model the VdW potential as a spherically symmetric barrier described later in Section \ref{Rezultki}. The magnetic dipole - dipole interaction potential $V_{dd}(\Br_{1}-\Br_{2})$ can be expressed in the following form:

\be \begin{aligned} V_{dd}= \frac{\mu_{0}}{4\pi\left | \Br_{1} - \Br_{2} \right |^{3}}& \left[ \boldsymbol{\mu}_{1} \cdot \boldsymbol{\mu}_{2}- \right.\\ &\left.+3 \left( \boldsymbol{\mu}_{1} \cdot \mathbf{n} \right)  \left( \boldsymbol{\mu}_{2} \cdot \mathbf{n} \right) \right] \end{aligned} \label{potencjaldipdip}\ee

where $\mathbf{n}=\frac{\Br_{1}-\Br_{2}}{\left | \Br_{1} - \Br_{2} \right |}$, $\mu_{0}$ stands for the vacuum magnetic permeability, $\boldsymbol{\mu}_{1}$ and $\boldsymbol{\mu}_{2}$ are magnetic moments of the two atoms with $|\boldsymbol{\mu}_{1}|=|\boldsymbol{\mu}_{2}|$. A magnetic moment of an arbitrary atom is connected with the total angular momentum of its unpaired electrons by:

\be\boldsymbol{\mu}=\mu_{B}g_{j}\BF \label{momentprzezspiny}\ee

where $\mu_{B}$ indicates the Bohr magneton, $g_{j}$ is the Land\'{e} g - factor and $\BF$ is the spin vector. Here we neglect the magnetic moment of the nucleus, which is smaller than the magnetic moment of the unpaired electrons by several orders of magnitude. Using \reff{momentprzezspiny} one obtains:

\be \begin{aligned} V_{dd}= \frac{\mu_{0}(\mu_{B}g_{j})^{2}}{4\pi\left | \Br_{1} - \Br_{2} \right |^{3}}& \left[ \BF_{1}\cdot \BF_{2}- \right.\\&\left.+3 \left( \BF_{1} \cdot \mathbf{n} \right)  \left( \BF_{2} \cdot \mathbf{n} \right) \right] \label{potencjaldipdip1} \end{aligned} \ee

This Hamiltonian may be divided into two parts, a center of mass part and a relative part i.e. $H=H_{CM}+H_{Rel}$ with:
\be
\begin{aligned}
H_{CM}&=-\frac{1}{2}\nabla_{R}+\frac{1}{2}\BR^{2}\\
H_{rel}&=-\frac{1}{2}\nabla_{r}+\frac{1}{2}\Br^{2}+V_{VdW}(r) \\ & \hspace{4mm}+\frac{g_{dd}}{r^{3}}\left[ \BF_{1}\cdot \BF_{2}-3 \left( \BF_{1} \cdot \mathbf{n} \right)  \left( \BF_{2} \cdot \mathbf{n} \right) \right]
\end{aligned}
\label{zmiennychrozdzielenie}
\ee

where $\BR=\frac{1}{\sqrt{2}}\left( \Br_{1}+\Br_{2} \right)$ is the center of mass coordinate and  $\Br=\frac{1}{\sqrt{2}}\left( \Br_{1}-\Br_{2} \right)$ stands for the relative motion coordinate~\footnote{Note somewhat unusual factor of $\sqrt{2}$ introduced here for symmetry}. The strength of the dipole - dipole interaction is characterized by the $g_{dd}$ constant which is equal to:
\be g_{dd}= \frac{\mu_{0}(\mu_{B}g_{j})^{2}}{8 \sqrt{2} \pi}\label{stalagdd}\ee

The eigenvalues of the $H_{CM}$ are simply that of the harmonic oscillator. In order to investigate the relative motion of the two atoms we observe that the total angular momentum is conserved:

\be\left[ \BJ,H_{rel} \right]\label{komutator}=0\ee 

where $\BJ$ stands for the total angular momentum of the system. The spherical symmetry of the system means that it is convenient to solve the relative motion problem in a total angular momentum basis. As we neglect a nuclear spin of an atom the total angular momentum operator is a sum of the total spin operator $\BF=\BF_{1}+\BF_{2}$ and the orbital momentum operator of the relative motion of the atoms $\BL$. Namely:

\be \mathbf{J}=\BF+\BL \label{totalnyoperator}\ee

Eigenfunction of the system in the chosen basis can be written as:

\be \begin{aligned}\Psi^{jm_{j}}_{n}(\Br)&=\sum_{l,f}a^{jm_{j}lf}_{n}\psi^{jm_{j}lf}_{n}\left( \Br \right)\\&=\sum_{l,f}a^{jm_{j}lf}_{n}\phi_{n}^{jlf}(r)\left|jm_{j}lf\right\rangle \\&=\sum_{l,f}a^{jm_{j}lf}_{n}\phi_{n}^{jlf}(r)\sum_{\substack {m_{l},m_{f}\\m_{l}+m_{f}=m_{j}}}\\& \hspace{4mm} \times C^{jm_{j}}_{lm_{l}fm_{f}} \left|lfm_{l}m_{f}\right\rangle \end{aligned}\label{funkcjawlasna}\ee

Here $j$ denotes the total angular momentum quantum number and $m_{j}$ the magnetic total angular momentum number, $l$ and $m_{l}$ stands for the orbital momentum and the magnetic orbital momentum quantum numbers respectively. The total spin and its projection values are indicated by $f$ and $m_{f}$ and $C^{jm_{j}}_{lm_{l}fm_{f}}$ denotes Clebsch - Gordan coefficients~\cite{Abramowitz1974}. Eigenfunctions are enumerated by the $n=0,1,...$ number and $a^{jm_{j}lf}_{n}$ indicate  constant coefficients.\\
\indent Our goal is to derive the radial Shr\"{o}dinger equations for $\phi^{jlf}_{n}$ with given $j$, $l$, $f$. In order to achieve this it is convenient to rewrite the dipole - dipole interaction potential in terms of the ladder operators: 

\be \begin{aligned} V_{dd}&=\frac{g_{dd}}{r^{3}}\left[\frac{1}{2}\left(F_{1+}F_{2-}+F_{1-}F_{2+}\right)+F_{1z}F_{2z}+\right.\\& \hspace{4mm} -3\left(F_{1+}n_{-}+F_{1-}n_{+}+F_{1z}n_{z}\right) \\ & \hspace{4mm} \left. \times \left(F_{2+}n_{-}+F_{2-}n_{+}+F_{2z}n_{z}\right)\right]\end{aligned}\label{potencjaldrabinkowy}\ee

with:

\be \begin{aligned} n_{+}&=\frac{x+iy}{2r}=-\sqrt{\frac{2\pi}{3}}Y^{1}_{1}(\theta,\varphi)\\ 
n_{-}&=\frac{x-iy}{2r}=\sqrt{\frac{2\pi}{3}}Y^{-1}_{1}(\theta,\varphi)\\
n_{z}&=\frac{z}{r}=\sqrt{\frac{4\pi}{3}}Y^{0}_{1}(\theta,\varphi)\\
F_{+}&=F_{x}+iF_{y} \\
F_{-}&=F_{x}-iF_{y} 
\end{aligned} \label{drabinkoweoperatory}\ee

Here $Y^{m_{l}}_{l}(\theta,\varphi)$ denotes a standard spherical harmonic in the spherical coordinates. We are now interested in the result of acting with the $V_{dd}$ operator on the single state $\psi^{jm_{j}lf}_{n}(\Br)$. Using \reff{potencjaldrabinkowy}, \reff{drabinkoweoperatory}, spin operators properties and the well - known formula for the product of two spherical harmonics (see for instance~\cite{productwolfram}) it can be shown that:

\be V_{dd}\psi^{jm_{j}lf}_{n}(\Br)=\frac{g_{dd}}{r^{3}}\sum_{l',f'}\alpha_{ll'ff'}\psi^{jm_{j}l'f'}_{n}(\Br)\label{dzialanie}\ee

with the following selection rules:

\be \begin{aligned} l'&=l+\Delta l \hspace{1in} & \Delta l&=0,\pm 2\\
f'&=f+\Delta f \hspace{1in} & \Delta f&=0,\pm 2
 \end{aligned} \label{regulywyboru}\ee
 
A value of the scalar coefficient $\alpha_{ll'ss'}$ is expressed by a product of Clebsh - Gordon coefficients determined by the angular momentum algebra.\\
\indent The preceding result might be understood by the fact that the dipole - dipole interaction operator is symmetric with respect to the exchange of the two particles. Thus it does not change a symmetry of the given $\psi^{jm_{j}lf}_{n}(\Br)$. Knowing \reff{dzialanie} we are able to find the radial Schr\"{o}dinger equation for the $\chi_{n}^{jlf}(r)\equiv r\phi_{n}^{jlf}(r)$ by the direct calculation:
 
 \be \begin{aligned} -&\frac{1}{2}\frac{d^{2}}{dr^{2}}\chi_{n}^{jlf}(r)+\frac{1}{2}r^{2}\chi_{n}^{jlf}(r)+\frac{l(l+1)}{2r^{2}}\chi_{n}^{jlf}(r)\\&+\frac{g_{dd}}{r^{3}}\sum_{l',f'}\alpha_{ll'ff'}\chi^{jl'f'}_{n}(r)=E^{j}_{n}\chi_{n}^{jlf}(r)\end{aligned}\label{rownaniarad}\ee
  
where $E^{j}_{n}$ is an eigenvalue and the short range potential $V_{VdW}(r)$ is incorporated in the boundary conditions (see Section \ref{Rezultki}). Here $\sqrt{\frac{\hbar^5}{m^3\omega}}$ is an unit of the $g_{dd}$ in the harmonic oscillator units.\\
\indent As can be seen in \reff{rownaniarad} in order to find a $\chi_{n}^{jlf}(r)$ one has to solve a system of the radial Schr\"{o}dinger equations for a fixed total angular momentum number $j$. Note that the number of equations in the system is determined by the maximum value of the total spin: $f_{max}=f_{1}+f_{2}$.
\section{Results} \label{Rezultki}

\begin{figure*}[]
\begin{centering}
\includegraphics[width=\textwidth]{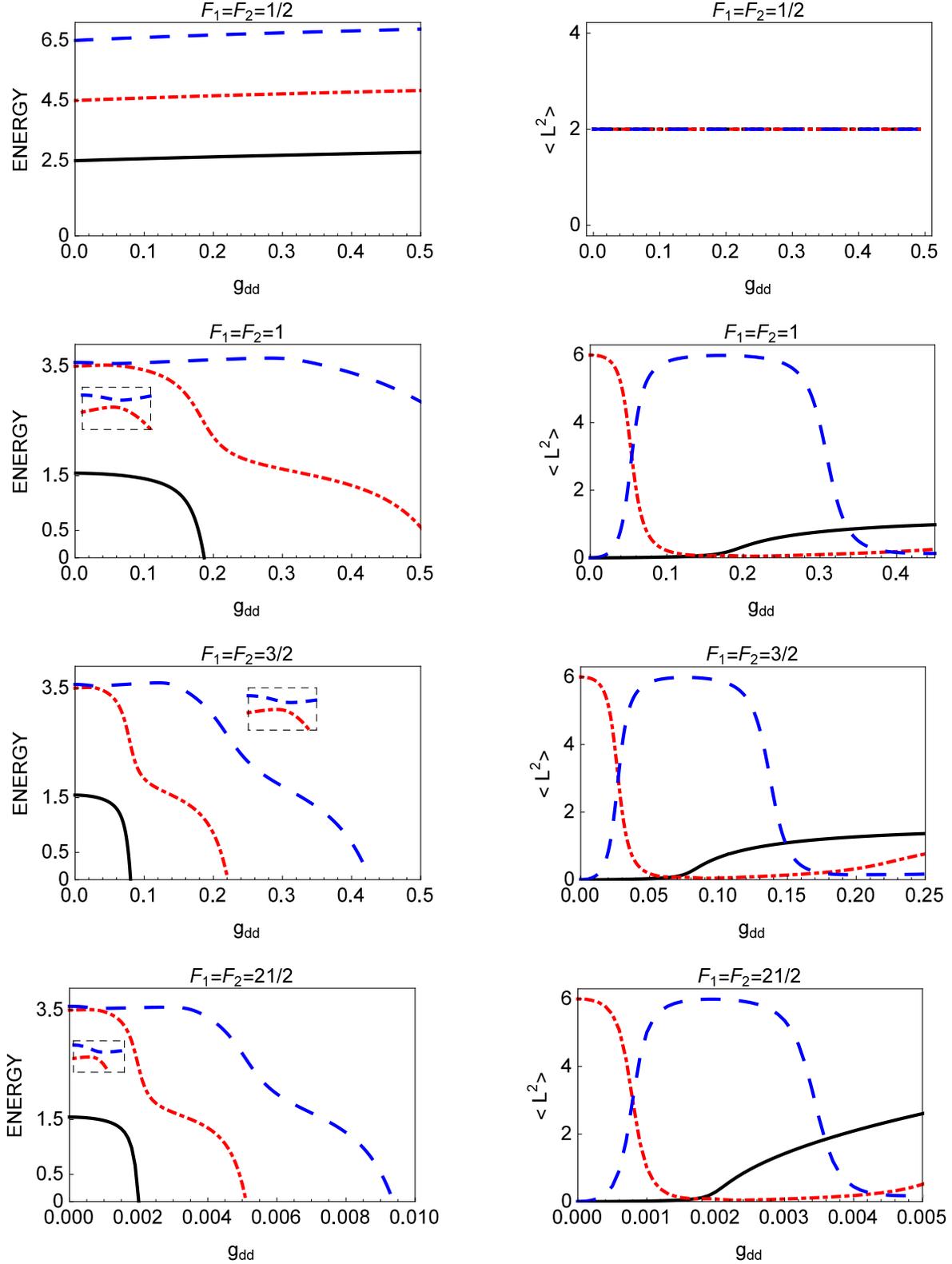}
\par\end{centering}
\caption{(Color on-line) Energy $E_{n}^{0}$ vs $g_{dd}$ and expected value of orbital angular momentum operator $\left < L^2 \right>$ for the $n=0,\hspace{1mm}1,\hspace{1mm}2$ and atoms of spin $f_{1}=f_{2}=\frac{1}{2},\hspace{1mm}1,\hspace{1mm}\frac{3}{2},\hspace{1mm}\frac{21}{2}$. The black solid line represents the ground state, the red dashed dotted line and blue dashed line indicate first and second excited states respectively. The insets magnify the anti - crossing area. Note different horizontal scale for $f_{1}=f_{2}=\frac{21}{2}$}
\label{enodgdd} 
\end{figure*}

We are interested in solving the system of the radial Schr\"{o}dinger equations introduced in Section \ref{Theomod}, in particular for the total angular momentum $j=0$. In order to accomplish this task we remind that the interaction potential $V(\Br_{1}-\Br_{2})$ introduced in section \ref{Theomod} consists of the short range Van der Waals potential also. We propose a simple model of such potential in the following form:

\be V_{VdW}=\left\{ \begin{aligned} 0 &\enspace &\text{for} \enspace r > b=100 \hspace{1mm} r_{0} \\ 
\infty&\enspace &\text{for} \enspace r \leq b=100 \hspace{1mm} r_{0}
\end{aligned} \right.\label{oddzialkontakt}\ee

where $r_{0}$ is the Bohr radius. We motivate our choice by the fact that the scattering length $a_{0}$ for a scattering process of a single particle on an infinite spherically symmetric potential barrier is equal to the radius of barrier i.e. $b=a_{0}$. A value of $b$ is determined by the numerical calculations for the dysprosium atoms~\cite{Petrov2012}.\\
\indent From the angular momentum algebra we also deduced that for eigenstate with $j=0$ the total spin number is equal to the orbital quantum number i.e. $l=f$. Thus for such states the corresponding coefficient matrix $\alpha_{ll'ff'}$ reduces to the $\alpha_{ll'}$ matrix. It can be also proved that $a_{00}=0$ for the arbitrary chosen $f_{1}=f_{2}$. We introduce the $\boldsymbol{\alpha}_{f_{1}}^{p}$ matrices of the $\alpha_{ll'}$ coefficients where $f_{1}$ is the single atom spin and $p=e,o$ indicates even or odd parity of the $l\text{ and }l'$. We calculate them for the various atomic spin values i.e. $f_{1}=f_{2}=\frac{1}{2},1,\frac{3}{2} \text{ and } \frac{21}{2}$. Our results can be found in the  Appendix \ref{wspolczynniki}.\\
\indent Knowledge of the $\alpha_{ll'ff'}$ coefficients allows us to solve numerically the system of the radial Schr\"{o}dinger equations of the form presented in \reff{rownaniarad}. We used the multi-parameter shooting method. We set the $b=0.04$ in the harmonic oscillator units which corresponds for the dysprosium-like atoms at the trap frequency $\omega \approx 2\pi \hspace{1mm}  3.2 \hspace{2mm} \text{kHz}$. For such a trap frequency the $g_{dd}=0.0006$ in the harmonic oscillator units. Our system admits two control parameters that may be changed by experimenters. Note that the $g_{dd}$ in the harmonic oscillator units depends on the trap frequency as $\sqrt{\omega}$, so it is tunable. One may also change the scattering length $a_{0}$ by the optical Feshbach resonances~\cite{Fedichev1996,Fatemi2000,Thalhammer2005,Blatt2011}, so that the $b$ value in the harmonic oscillator units may be kept constant while one changes the trap frequency.\\
\indent In Fig. \ref{enodgdd} we present the eigenvalues $E^{0}_{n}$ with $n=0,1,2$ as a function of $g_{dd}$ for  atoms with different spins. For atoms with the spin $f_{1}=f_{2}=1,\hspace{1mm}\frac{3}{2}\hspace{1mm} \text{and} \hspace{1mm} \frac{21}{2}$ we consider only solutions for the even orbital angular momentum quantum number $l$. In the case of odd $l$ results are qualitatively the same.\\
\indent For spin $\frac{1}{2}$ atoms the energy values rise very slowly as $g_{dd}$ rises. The radial part of $\psi^{0011}_{n}(\Br)$ is simply the $\phi_{n}^{011}(r)$, so the expected value of the orbital angular momentum operator $\left < L^{2} \right>$ is constant and equal $\left < L^{2} \right>=2$ for all $n$.\\
\indent For the higher spin values we observe more complex behaviour. First of all, the energy values for $n=0,\hspace{1mm}1 \hspace{1mm} \text{and} \hspace{1mm} 2$ are highly dependent on the value of $g_{dd}$. For low values of $g_{dd}$ eigenvalues vary slightly, then for higher values they decrease rapidly. We observe also the presence of anti - crossings between consecutive lines $E^{0}_{n}(g_{dd})$ accompanied by changes of the $\left < L^{2} \right>$. This is due to changes in the structure of the radial part of eigenstates. From the \reff{rownaniarad} we notice that the radial part of eigenstate is a linear combination of the $\phi_{n}^{0ll}(r)$ where in this case $l \in \left\lbrace 0,2,..,2\cdot f_{1} \right\rbrace$. As the $g_{dd}$ rises the weight of each $\phi_{n}^{0ll}(r)$ function varies i.e. values of $a^{00ll}_{n}$ coefficient varies. For instance, we see that for low $g_{dd}$ the ground state consists of almost only the s - state ($\phi_{0}^{000}(r)$), whereas as we increase the trap frequency, contributions of the functions with higher $l$ grow. The ground state starts to ''rotate''. This feature resembles the Einstein - de Haas effect~\cite{Einstein1915}, although it is caused only by the internal spin - spin interactions between two atoms without any influence of external fields.\\
\indent Fig. \ref{enodgdd} also illustrates that the bigger atomic spin is, the lower trap frequency is needed to observe above effects. In addition, the effect of changes in the expected value of orbital angular momentum is stronger for larger atomic spin values. It seems that at least it is possible to check our model experimentally using the system of the dysprosium atoms with the $\frac{21}{2}$ spin.\\
\begin{figure*}[]
\begin{centering}
\includegraphics[width=0.9\textwidth]{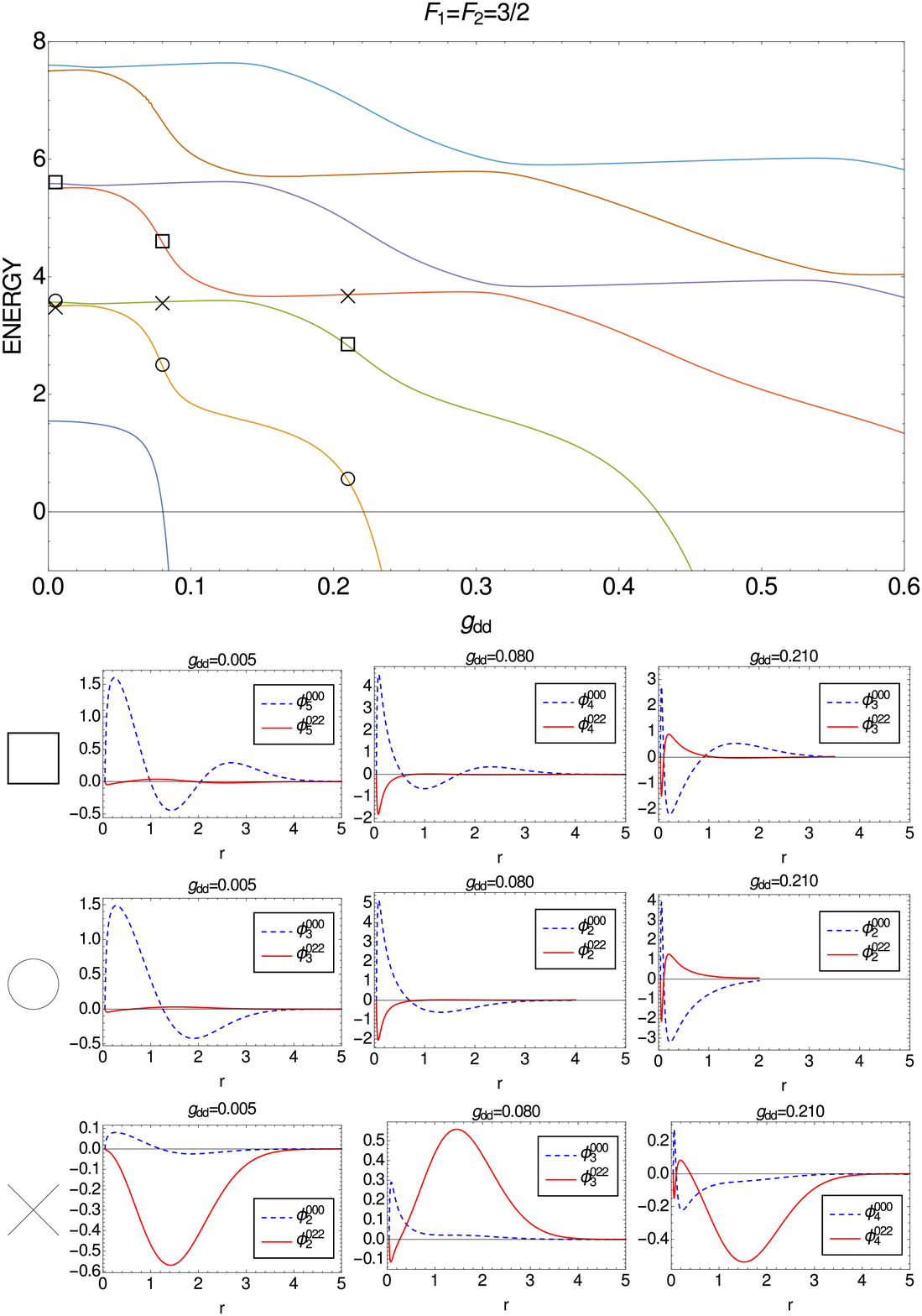}
\par\end{centering}
\caption{(Color on-line) Composition of eigenstates for different eigenvalues $E_{n}^{0}$ for the $\frac{3}{2}$ spin atoms. The blue dashed lines indicate radial functions $\phi_{n}^{000}(r)$ with the orbital quantum number $l=0$ for the given value of $g_{dd}$ vs radial variable $r$ and the red solid lines represent radial functions $\phi_{n}^{022}$ for the given $g_{dd}$ with the orbital quantum number $l=2$ vs radial variable $r$. Square, circle and cross stay for sets of eigenstates with the same composition of $\phi_{n}^{0ll}(r)$ functions.}
\label{landzen} 
\end{figure*}
\indent The nature of anti - crossings in Fig. \ref{enodgdd} can be explained by Landau - Zener theory~\cite{Landau,Zener1932,stuel1932,Majorana} as depicted in Fig. \ref{landzen}. As an example we used $\frac{3}{2}$ spin atoms. A composition of the eigenstate corresponding to the eigenvalue $E_{n}(g_{dd})$ is not conserved 
along given energy line, but it propagates along straight lines upward or downward.\\
\indent Motivated by experiments under development~\cite{Maier2014,Tang2015,Kadau2015} we based our calculations on  dysprosium parameters. Our model of the dipole - dipole interactions between two atoms reveals a non-trivial dependence of two atoms in a harmonic trap system on  the trap frequency. We showed that increasing $\omega$ the system undergoes an analog of Einstein - de Hass effect. Such a behaviour is a result of spin - spin interaction and its coupling to the orbital angular momentum. We have also found the Landau - Zener anti - crossings in the energy levels of the system. Our results may be checked experimentally for the dysprosium atoms. Of course, proposed model is oversimplified in this case as dysprosium atoms are not exactly spherically symmetric~\cite{Petrov2012}.
\begin{acknowledgments}
The authors acknowledge fruitful conversations with Mariusz Gajda. This
work was supported by the (Polish) National Science Center Grant No. DEC-
2012/04/A/ST2/00090.
\end{acknowledgments}

\appendix
\section{The $\alpha_{ll'ff'}$ coefficients}\label{wspolczynniki}

\paragraph{Spin $\frac{1}{2}$}

For the spin $\frac{1}{2}$ particles $\boldsymbol{\alpha}_{\frac{1}{2}}^{p}$ are scalars. For the singlet state we obtain:

\be \boldsymbol{\alpha}^{e}_{\frac{1}{2}}=0 \label{signletalfa}\ee

which means that the singlet state is not affected by the dipole - dipole interactions. The corresponding matrix for the triplet states is:

\be \boldsymbol{\alpha}^{o}_{\frac{1}{2}}=1 \label{tripletalfa}\ee

and the dipole - dipole interaction is repulsive. The above results allow one to investigate the contact interactions between the two $\frac{1}{2}$ spin atoms and the dipole - dipole interactions in parallel.\\
For $j=1$ the only non vanishing coefficient is $\alpha_{1111}$ equal to:

\be \alpha_{1111}=-\frac{1}{2} \label{tripletalfaj1}\ee

thus in this case the dipole - dipole interaction is attractive.

\paragraph{Spin $1$}

The coefficient matrix for the even orbital angular momentum quantum numbers $l,l'=0,2$ can be written as:

\be \boldsymbol{\alpha}^{e}_{1}=\left( \begin{array}{cc}
0 & \sqrt{2} \\
\sqrt{2} & 2
\end{array} \right) \label{spin1l02} \ee
For the $l,l'=1$ we obtain:
\be\boldsymbol{\alpha}_{1}^{o}=2 \label{spin1l1}\ee

\paragraph{Spin $\frac{3}{2}$}
The coefficient matrix for the even orbital angular momentum quantum numbers $l,l'=0,2$ can be expressed by:
\be\boldsymbol{\alpha}_{\frac{3}{2}}^{e}=\left( \begin{array}{cc}
0 & 3 \\
3 & 3
\end{array} \right)\label{spin3/2l02}\ee
For the $l,l'=1,3$ we obtain:
\be\boldsymbol{\alpha}_{\frac{3}{2}}^{o}=\left( \begin{array}{cc}
\frac{17}{5} & \frac{9}{5} \\
\frac{9}{5} & \frac{18}{5}
\label{spin3/2l13}
\end{array} \right)\ee

\paragraph{Spin $\frac{21}{2}$}

The coefficient matrix for the even orbital angular momentum quantum numbers $l,l'=\left\{ 0,2,...,18,20\right\}$ can be written as:

\begin{widetext}

 \be \resizebox{0.9\textwidth}{!} { $ \boldsymbol{\alpha}_{\frac{21}{2}}^{e}=\left( \begin{array}{*{11}{c}}  0 & 6\sqrt{322} & 0 & 0 & 0 & 0 & 0 & 0 & 0 & 0 & 0  \\
6\sqrt{322} & \frac{489}{7} & \frac{18\sqrt{1235}}{7} & 0 & 0 & 0 & 0 & 0 & 0 & 0 & 0 \\
0 & \frac{18\sqrt{1235}}{7} & \frac{5030}{77} & \frac{180\sqrt{357}}{11\sqrt{13}} & 0 & 0 & 0 & 0 & 0 & 0 & 0 \\
0 & 0 & \frac{180\sqrt{357}}{11\sqrt{13}} & \frac{735}{11} & \frac{84\sqrt{203}}{\sqrt{221}} & 0 & 0 & 0 & 0 & 0 & 0 \\
0 & 0 & 0 & \frac{84\sqrt{203}}{\sqrt{221}} & \frac{1332}{19} & \frac{540\sqrt{806}}{19\sqrt{119}} & 0 & 0 & 0 & 0 & 0 \\
0 & 0 & 0 & 0 & \frac{540\sqrt{806}}{19\sqrt{119}} & \frac{32615}{437} & \frac{2178\sqrt{17}}{23\sqrt{35}} & 0 & 0 & 0 & 0 \\
0 & 0 & 0 & 0 & 0 & \frac{2178\sqrt{17}}{23\sqrt{35}} & \frac{1846}{23} & \frac{182\sqrt{14}}{\sqrt{145}} & 0 & 0 & 0 \\
0 & 0 & 0 & 0 & 0 & 0 & \frac{182\sqrt{14}}{\sqrt{145}} & \frac{2695}{11} & \frac{360\sqrt{4921}}{31\sqrt{319}} & 0 & 0 \\
0 & 0 & 0 & 0 & 0 & 0 & 0 & \frac{360\sqrt{4921}}{31\sqrt{319}}  & \frac{20536}{217} & \frac{918\sqrt{26}}{7\sqrt{407}} & 0 \\
0 & 0 & 0 & 0 & 0 & 0 & 0 & 0 & \frac{918\sqrt{26}}{7\sqrt{407}} & \frac{9405}{91} & \frac{570\sqrt{7}}{13\sqrt{37}} \\
0 & 0 & 0 & 0 & 0 & 0 & 0 & 0 & 0 & \frac{570\sqrt{7}}{13\sqrt{37}} & \frac{1470}{13}
\end{array} \right) $ } \label{spin21/2l0220}\ee
\end{widetext}
Note that the above is the tri-diagonal band matrix as was noted in \reff{regulywyboru}.

\bibliographystyle{apsrev4-1}

\bibliography{articlebib}

\end{document}